\documentclass[prc,twocolumn,showpacs,amsmath,amssymb,nofootinbib,superscriptaddress]{revtex4}
\usepackage{mathrsfs}
\usepackage{amsfonts}
\usepackage{mathptmx}
\usepackage{epsfig}
\usepackage{graphicx}
\usepackage{slashed}
\usepackage{color}

\def\comment#1{}

\begin{document}

\title[]{Surface tension of the core-crust interface of neutron stars with global charge neutrality}

\author{Jorge A. Rueda}
\affiliation{Dipartimento di Fisica and ICRA, Sapienza Universit\`a
di Roma, P.le Aldo Moro 5, I-00185 Rome, Italy}%
\affiliation{ICRANet, P.zza della Repubblica 10, I-65122 Pescara,
Italy}%
\author{Remo Ruffini}
\affiliation{Dipartimento di Fisica and ICRA, Sapienza Universit\`a
di Roma, P.le Aldo Moro 5, I-00185 Rome, Italy}%
\affiliation{ICRANet, P.zza della Repubblica 10, I-65122 Pescara,
Italy}%
\affiliation{ICRANet, University of Nice-Sophia Antipolis, 28 Av. de
Valrose, 06103 Nice Cedex 2, France}
\author{Yuan-Bin Wu}
\affiliation{Dipartimento di Fisica and ICRA, Sapienza Universit\`a
di Roma, P.le Aldo Moro 5, I-00185 Rome, Italy}%
\affiliation{ICRANet, P.zza della Repubblica 10, I-65122 Pescara,
Italy}%
\affiliation{ICRANet, University of Nice-Sophia Antipolis, 28 Av. de
Valrose, 06103 Nice Cedex 2, France}
\author{She-Sheng Xue}
\affiliation{Dipartimento di Fisica and ICRA, Sapienza Universit\`a
di Roma, P.le Aldo Moro 5, I-00185 Rome, Italy}%
\affiliation{ICRANet, P.zza della Repubblica 10, I-65122 Pescara,
Italy}%

\begin{abstract}
It has been shown recently that taking into account strong, weak,
electromagnetic, and gravitational interactions, and fulfilling the
global charge neutrality of the system, a transition layer will
happen between the core and crust of neutron stars, at the nuclear
saturation density. We use relativistic mean field theory together
with the Thomas-Fermi approximation to study the detailed structure
of this transition layer and calculate its surface and Coulomb
energy. We find that the surface tension is proportional to a
power-law function of the baryon number density in the core bulk
region. We also analyze the influence of the electron component and
the gravitational field on the structure of the transition layer and
the value of the surface tension to compare and contrast with known
phenomenological results in nuclear physics. Based on the above
results we study the instability against Bohr-Wheeler surface
deformations in the case of neutron stars obeying global charge
neutrality. Assuming the core-crust transition at nuclear density
$\rho_{core}\approx 2.7\times 10^{14}$ g cm$^{-3}$, we find that the
instability sets the upper limit to the crust density,
$\rho_{crust}^{crit}\approx 1.2 \times 10^{14}$ g cm$^{-3}$. This
result implies a nonzero lower limit to the maximum electric field
of the core-crust transition surface and makes inaccessible a limit
of quasilocal charge neutrality in the limit
$\rho_{crust}=\rho_{core}$. The general framework presented here can
be also applied to study the stability of sharp phase transitions in
hybrid stars as well as in strange stars, both bare and with outer
crust. The results of this work open the way to a more general
analysis of the stability of these transition surfaces, accounting
for other effects such as gravitational binding, centrifugal
repulsion, magnetic field induced by rotating electric field, and
therefore magnetic dipole-dipole interactions.
\end{abstract}

\pacs{26.60.-c, 97.60.Jd, 04.20.-q, 04.40.Dg}


\maketitle

\section{Introduction}
\label{sec:intro}

The relativistic mean field theory (RMFT) of nuclear matter and the
Thomas-Fermi model have attracted great attention during the last
few decades. The simplest relativistic model of nuclear matter that
accounts for the saturation properties of symmetric nuclear matter
includes one scalar field which gives the attractive long-range part
of the nuclear force and one vector field which gives the repulsive
short-range; these two meson fields interact with nucleons through
Yukawa couplings. This so-called $\sigma$-$\omega$ model has been
considered by Duerr \cite{Duerr1}, Miller and Green \cite{Miller1},
and later by Walecka \cite{Walecka1}. The physical understanding of
this model has been very well studied the literature
\cite{Boguta1,Boguta2,Lee1,Lee2,Lee3,Boguta3,Boguta4,Boguta5}. As
recognized in Ref.~\cite{Boguta2}, it is necessary to introduce
additional isovector fields to obtain agreement with the empirical
symmetry energy of nuclear matter at the saturation density. The
model, containing Dirac nucleons together with a self-interacting
scalar $\sigma$ and a vector meson $\omega$ as well as an isovector
meson $\rho$, has been widely used to this end.

With a very limited number of parameters, the RMFT has been shown to
be able to give a quantitative description of a variety of nuclear
properties \cite{Serot1,Ring1,Bender1}. Recently, taking into
account the electromagnetic and weak interactions, the RMFT with the
Thomas-Fermi approximation has gained remarkable successes in
understanding the inhomogeneous structures and properties of
low-density nuclear matter which is realized in supernovae cores or
in the crusts of neutron stars (see, e.g.,
Refs.~\cite{Maruyama1,Avancini1,Okamoto1,Grill1}). The surface
properties of nuclear matter such as surface tension and curvature
energy play an important role in the description of these structures
and also in other phenomena, for instance saddle-point
configurations in nuclear fission, fragment distributions in
heavy-ion collisions, and phase transition between different phases
of nuclear matter.

The nuclear surface properties at saturation density have been
analyzed for a long time in the semi-infinite nuclear matter model
using RMFT \cite{Walecka1} or effective field theory
\cite{Furnstahl1,Furnstahl2,Serot2} with the Thomas-Fermi
approximation or Hartree-Fock approximation
\cite{Boguta2,Brack1,Sharma1,Eiff1,Eiff2,Eiff3,Centelles1,Estal2,Patra1,Danielewicz1}.
In the supranuclear regime realized in the interior of neutron
stars, there is the possibility that  phase transition occurs from
hadronic to pion and kaon condensed phase as well as to quark matter
phase (see, e.g., \cite{Glendenning1,Glendenning2,Glendenning3}).
The surface tension of the transition layer between the hadronic and
kaon condensed or quark matter phases has been calculated in the
semi-infinite matter model, and the surface tension plays an
important role in the structure of the phase transition region
\cite{Christiansen1,Alford1}. In the low-density (density smaller
than the saturation density) case, as pointed out in
\cite{Ravenhall1}, the shape of constituent nuclei is expected to
change from spherical droplet to the so-called nuclear pasta
structures such as cylindrical rod, slab, cylindrical tube, and
spherical bubble. The surface tensions of nuclear pasta structures
have been investigated and it has been pointed out that the pasta
phase strongly depends on the value of the surface tension
\cite{Maruyama1,Avancini1,Grill1}.

The importance of the extension of the Thomas-Fermi approximation to
general relativistic systems such as neutron stars was emphasized in
Ref.~\cite{RuedaPLB}. We showed there that the traditionally imposed
condition of local charge neutrality is not consistent with the
field equations and microphysical equilibrium for a system of
neutrons, protons, and electrons in $\beta$ equilibrium and obeying
relativistic quantum statistics. Thus, only the condition of global
but not local charge neutrality can be imposed. This leads to the
appearance of gravito-polarization in the cores of neutron stars.
The generalization of such a work to the case where the strong
interactions between nucleons are accounted for was presented in
\cite{Rueda1}. Both the Thomas-Fermi approximation and RMFT were
used. It was shown that the Einstein-Maxwell-Thomas-Fermi (EMTF)
system of equations within RMFT supersede the traditional
Tolman-Oppenheimer-Volkoff (TOV) \cite{Tolman1,Oppenheimer1}
equations used for the construction of neutron star configurations.

Realistic neutron star configurations including all the interactions
between particles and the presence of a crust below nuclear density,
were constructed in Ref.~\cite{Belvedere1} by solving numerically
the EMTF equations fulfilling the condition of global charge
neutrality. As pointed out in \cite{Belvedere1}, the self-consistent
solution of these new equations of equilibrium leads to the
existence of a transition layer between the core and the crust of
the star. This is markedly different from the neutron star structure
obtained from the solution of the TOV equations imposing local
charge neutrality (see e.g., \cite{Haensel1}), leading to a new
mass-radius ($M$-$R$) relation of neutron stars. The core-crust
transition layer in our configurations occurs near the nuclear
saturation density $\rho_{nucl}$. The core (bulk region) inside this
transition layer is a hadronic phase, and the crust outside this
transition is composed of the nuclei lattice and the ocean of
relativistic degenerate electrons and possibly neutrons at densities
below nuclear saturation and larger than the estimated neutron drip
value $\sim 4.3\times 10^{11}$ g cm$^{-3}$. Inside the transition
region a very strong electric field which is overwhelming the
critical value $E_c=m^2_e c^3/(e \hbar)$ for vacuum breakdown is
developed, where $m_e$ is the electron rest mass. The $e^+e^-$ pair
creation from vacuum is, however, forbidden in the system due to the
Pauli blocking of degenerate electrons.

In this article we study the surface properties of this transition
layer formed near the nuclear saturation density. We calculate all
the contributions to the surface tension as well as the
electrostatic energy stored in this core-crust layer. We analyze the
stability of these systems under the Bohr-Wheeler fission mechanism
\cite{Bohr1}. We analyze the role of the influence of the
gravitational field on the structure of the transition layer and the
surface tension. We also compare and contrast the surface energy of
these neutron stars with the phenomenological results in nuclear
physics.

The article is organized as follows. In Sec.~\ref{sec:stw}, we
present the general formulation of the surface tension as well as
the Coulomb energy for this core-crust transition layer. We
formulate in Sec.~\ref{sec:equs} the relativistic equations for a
system of neutrons, protons, and electrons fulfilling the strong,
electromagnetic, and gravitational interactions as well as $\beta$
equilibrium. In Sec.~\ref{sec:SIM}, we use the semi-infinite matter
model \cite{Baym1} to formulate the equations governing the surface
tension for the transition layer of this system when the electron
density is nearly equal to the proton density in the core bulk
region. In Sec.~\ref{sec:stns}, we study the surface tension and the
Coulomb energy, neglecting the presence of the crust and the
gravitational interaction. We calculate the surface structure and
solve these equations to obtain the surface tension and the Coulomb
energy at the nuclear saturation density in Sec.\ref{sec:stnns}.
Then we study in Sec.~\ref{sec:dbn} the dependence of the surface
tension and the Coulomb energy on the baryon number density. In
Sec.~\ref{sec:dne}, we study the influence of fermion densities in
the outside region (crust) on the surface tension and the Coulomb
energy. In Sec.~\ref{sec:stwg}, we study the structure and the
surface tension as well as the Coulomb energy for the core-crust
transition region in the presence of the gravitational field. We
finally summarize and conclude in Sec.~\ref{sec:sum}. We use units
with $\hbar = c = 1$ throughout the article.

\section{Relativistic equations of motion and surface tension}
\label{sec:stw}

\subsection{Relativistic equations of motion}
\label{sec:equs}

As described in Ref.~\cite{Belvedere1}, the system we consider is
composed of degenerate neutrons, protons, and electrons fulfilling
global charge neutrality and $\beta$ equilibrium. We include the
strong, electromagnetic, weak, and gravitational interactions. To
describe the nuclear interactions, we employ the RMFT with the
Thomas-Fermi approximation. We adopt the phenomenological nuclear
model of Boguta and Bodmer \cite{Boguta2}.

We introduce the nonrotating spherically symmetric spacetime metric
\begin{equation} \label{ssmetric}
  ds^2 = {\rm{e}}^{\nu(r)} dt^2 - {\rm{e}}^{\lambda(r)} dr^2 - r^2 d\theta^2
  - r^2 \sin^2\theta d \varphi^2,
\end{equation}
where the $\nu(r)$ and $\lambda(r)$ are only functions of the radial
coordinate $r$.

Within the Thomas-Fermi approximation and mean-field approximation,
we can obtain the full system of general relativistic equations. A
detailed description of this model can be found in
Ref.~\cite{Belvedere1}. We are here interested in the core-crust
transition layer, which as we have shown happens in a tiny region
\cite{Belvedere1} with a characteristic length scale $\sim
\lambda_{e} = \hbar /(m_e c) \sim 100$ fm. Correspondingly, the
metric functions are essentially constant in this region. Thus in
the core-crust transition layer the system of equations can be
written as
\begin{eqnarray}
  & & \frac{d^2 V}{dr^2} + \frac{2}{r} \frac{dV}{dr}
  = -4\pi e {\rm{e}}^{\nu_{core}/2} {\rm{e}}^{\lambda_{core}} (n_p -
  n_e), \label{eqcomg}\\
  & & \frac{d^2 \sigma}{dr^2} + \frac{2}{r} \frac{d \sigma}{dr} =
  {\rm{e}}^{\lambda_{core}} [\partial_{\sigma} U(\sigma) + g_s n_s],
  \label{eqsigg} \\
  & & \frac{d^2 \omega}{dr^2} + \frac{2}{r} \frac{d \omega}{dr} =
  -{\rm{e}}^{\lambda_{core}} (g_{\omega} J_0^{\omega} - m_{\omega}^2 \omega),
  \label{eqomg} \\
  & & \frac{d^2 \rho}{dr^2} + \frac{2}{r} \frac{d \rho}{dr} =
  -{\rm{e}}^{\lambda_{core}} (g_{\rho} J_0^{\rho} - m_{\rho}^2 \rho),
  \label{eqrhog}\\
  & & E_e^{F} = {\rm{e}}^{\nu_{core}/2}\mu_e - e V = {\rm{constant}}, \label{eqeg}\\
  & & E_p^{F} = {\rm{e}}^{\nu_{core}/2}\mu_p + g_{\omega} \omega + g_{\rho} \rho
  + e V = {\rm{constant}}, \label{eqpg}\\
  & & E_n^{F} = {\rm{e}}^{\nu_{core}/2}\mu_n + g_{\omega} \omega - g_{\rho} \rho  =
  {\rm{constant}}, \label{eqng}
\end{eqnarray}
where the notation $\omega_0 \equiv \omega$, $\rho_0 \equiv \rho$,
and $A_0 \equiv V$ for the time components of the meson fields have
been introduced. Here $\mu_i = \sqrt{(P_i^F)^2 + \tilde{m}_i^2}$ and
$n_i = (P_i^F)^3/(3\pi^2)$ are the free chemical potential and
number density of the $i$-fermion species with Fermi momentum
$P_i^F$. The particle effective masses are $\tilde{m}_N = m_N + g_s
\sigma$ and $\tilde{m}_e = m_e$, where $m_i$ stands for the rest
mass of each $i$-fermion species. ${\rm{e}}^{\nu_{core}}\equiv
{\rm{e}}^{\nu(R_{core})}$ and ${\rm{e}}^{\lambda_{core}}\equiv
{\rm{e}}^{\lambda(R_{core})}$ are the metric functions evaluated at
the core radius $R_{core}$. $g_s$, $g_{\omega}$, and $g_{\rho}$ are
the coupling constants of the $\sigma$, $\omega$ and $\rho$ fields,
$e$ is the fundamental electric charge, and $m_{\omega}$, and
$m_{\rho}$ are the masses of $\omega$ and $\rho$. The scalar
self-interaction potential is
\begin{equation}
  U(\sigma) = \frac{1}{2} m_{\sigma}^2 \sigma^2 + \frac{1}{3} g_2
  \sigma^3 + \frac{1}{4} g_3 \sigma^4,
\end{equation}
with the $\sigma$ meson mass $m_{\sigma}$ and the third- and fourth-
order constants of the self-scalar interactions $g_2$ and $g_3$.

The generalized Fermi energies of electrons, protons, and neutrons,
$E_e^F$, $E_p^F$, and $E_n^F$ (so-called the Klein potentials
\cite{Rueda1}), are linked by the $\beta$ equilibrium \cite{Boguta6}
of protons, neutrons, and electrons,
\begin{equation} \label{betaeq}
  E_n^F = E_p^F + E_e^F.
\end{equation}

The scalar density $n_s$ is given by the expectation value
\begin{equation} \label{nsden}
  n_s = \langle \bar{\psi}_N \psi_N\rangle = \frac{2}{(2\pi)^3}
  \sum_{i=n,p} \int_0^{P_i^F} d^3 k \frac{\tilde{m}_N}{\epsilon_i(k)},
\end{equation}
where $\epsilon_i(k) = \sqrt{k^2 + \tilde{m}^2_i}$ is the
single-particle energy, and $\psi_N$ is the nucleon isospin doublet.
In the static case, the nonvanishing components of the currents are
\begin{eqnarray}
  J_0^{ch} &=& {\rm{e}}^{\nu_{core}/2} (n_p - n_e),\\
  J_0^{\omega} &=& {\rm{e}}^{\nu_{core}/2} (n_n + n_p),\\
  J_0^{\rho} &=& {\rm{e}}^{\nu_{core}/2} (n_p -n_n),
\end{eqnarray}
here $n_b = n_p + n_n$ is the baryon number density.

The parameters of the nuclear model, namely the coupling constants
$g_{s}$ , $g_{\omega}$, and $g_{\rho}$, the meson masses
$m_{\sigma}$, $m_{\omega}$, and $m_{\rho}$, and the self-scalar
interaction constants $g_2$ and $g_3$ are fixed by fitting nuclear
experimental data. We here use the parameters of the NL3
parametrization \cite{Lalazissis1}, shown in Table \ref{tablenl3}.

\begin{table}[h]  \addtolength{\tabcolsep}{5pt}
\begin{center}
\begin{tabular}{lc|lc}
\hline  $m_{\sigma}$ (MeV) & $508.194$& $g_{\omega}$       & $12.8680$\\
        $m_{\omega}$ (MeV) & $782.501$& $g_{\rho}$         & $4.4740$\\
        $m_{\rho}$    (MeV) & $763.000$& $g_2$ (${\rm fm}^{-1}$)  & $-10.4310$\\
        $g_{s}$            & $10.2170$& $g_3$              & $-28.8850$\\
\hline
\end{tabular}
\end{center}
\caption{The parameters of the nuclear model from NL3.}
\label{tablenl3}
\end{table}

Since the equation of state (EOS) obtained from the RMFT is very
stiff (see, e.g., \cite{Haensel1}), it is natural to evaluate its
consequences on causality. In order to do this, we compute the
material sound velocity, $v^2_s= d \varepsilon/d {\cal P}$, as a
function of the central density $\rho(0)=\varepsilon(0)/c^2$ of the
configuration, where $\varepsilon=T^0_0$ and ${\cal P}=-T^1_1$ are
total energy-density and pressure of the system, $T^0_0$ and $T^1_1$
being the 0--0 and 1--1 components of the energy-momentum tensor
\cite{Belvedere1}.

\begin{figure}[h]
\includegraphics[width=0.95\columnwidth]{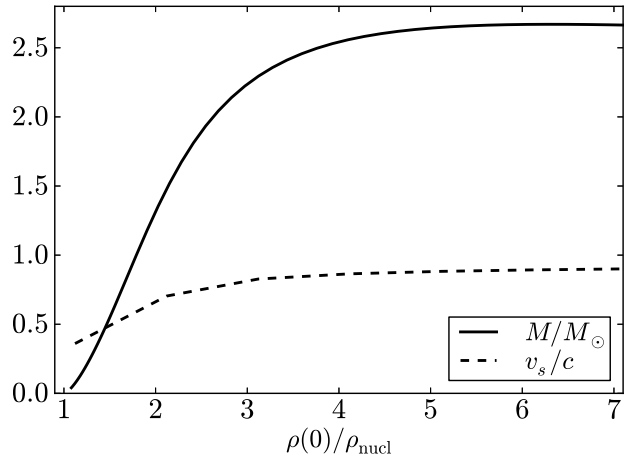}
\caption{The dependence of the total mass $M$ of the star and the
material sound velocity $v_s$ on the central density $\rho(0)$ of
the configuration.} \label{Mrhovs}
\end{figure}

The result is shown in Fig.~\ref{Mrhovs}. We recall that the
instability against gravitational collapse sets in at the turning
point in the $M$-$\rho(0)$ diagram, namely at the first maximum in
the sequence of equilibrium configurations with increasing central
density, namely $dM/d\rho(0)=0$. Such a point gives us the maximum
stable mass $M_{max}$, which for the present EOS is $M_{max}\approx
2.67~M_\odot$, where $M_\odot$ is the solar mass. It can be seen
from Fig.~\ref{Mrhovs} that $v_s < c$, where $c$ is the velocity of
light, at any density in the entire range of $\rho(0)$ of the stable
configurations, and therefore the used EOS does not violate
causality.

It is important to mention that the above critical point for the
gravitational collapse does not coincide with the point of
backbending of the $M$-$R$ relation (see, e.g., Figs.~6 and 14 in
\cite{Belvedere1}). Therefore the backbending in the $M$-$R$ diagram
does not indicate any sort of instability.

\subsection{Surface tension for semi-infinite matter}
\label{sec:SIM}

As shown in \cite{Belvedere1}, in the bulk hadronic phase of neutron
star cores, the charge separation is very small, so the electron
density $n_{eb}$ is nearly equal to the proton density $n_{pb}$. In
addition, the core-crust transition layer has a characteristic
length scale of the order of the electron Compton wavelength; this
is very small compared to the radius of neutron stars. So it is a
good approximation to use the semi-infinite matter model to
construct the surface tension for the system we consider here. We
construct the surface tension for the transition layer of this
system following the method of Baym, Bethe, and Pethick (BBP)
\cite{Baym1}.

In the semi-infinite matter model, one assumes a plane surface with
small thickness compared with the bulk region size separating two
semi-infinite regions, represented here by the inside core bulk and
the outside crust. The number density of the $i$-fermion species
($i=n,p,e$) $n_i(r)$ approaches the bulk density of the $i$-fermion
species $n_{ib}$ as the position $\bar{r}\equiv
(r-R_{core})\rightarrow -\infty$, and approaches the density in the
outside region of the $i$-fermion species $n_{io}$ as the $\bar{r}
\rightarrow +\infty$. To construct the surface tension, one imagines
a reference system with a sharp surface at $\bar{r}=a_i$ at which
fermion densities and meson fields fall discontinuously from the
core bulk region to the outside crust region. Following
Ref.~\cite{Baym1}, the location of the reference surface for the
$i$-fermion species is defined by the condition that the reference
system has the same number of the $i$-fermion species as the
original system. Following the definition of fermion number in the
curved spacetime, Eq.~(\ref{ssmetric}) (see, e.g., \cite{Lee4}), the
$i$-fermion species number $N_i$ is given by
\begin{equation}
  N_i = 4\pi \int {\rm e}^{\lambda/2} r^2 n_i(r) d r.
\end{equation}
Since the metric functions are constant in the surface region we
consider as described in Sec.~\ref{sec:equs}, and the size of the
surface region is very small compared to the radius of neutron
stars, we can treat ${\rm e}^{\lambda/2} r^2$ as a constant in the
integral, the location of the reference surface for the $i$-fermion
species is given by
\begin{equation} \label{sffg}
  \int_{-\infty}^{a_i} d \bar{r} [n_i(\bar{r}) - n_{ib}]
  + \int_{a_i}^{\infty} d r [n_i(\bar{r}) - n_{io}]
  = 0, \quad i = n, p , e.
\end{equation}
Applying the definition of the reference surface in Eq.~(\ref{sffg})
to the neutron, proton, and electron distributions yields slightly
different reference surfaces.

Similar to the definition of the reference surface for the fermion,
we define the location of the reference surfaces for meson fields by
\begin{equation} \label{sfmfg}
  \int_{-\infty}^{a_i} d \bar{r} [F_i(\bar{r}) - F_{ib}]
  + \int_{a_i}^{\infty} d \bar{r} [F_i(\bar{r}) - F_{io}]
  = 0, \quad i = \sigma, \omega, \rho,
\end{equation}
where $F_i(\vec{r})$ is the time component of the $i$-meson field,
$F_{ib}$ is the time component of the $i$-meson field in the bulk
region, and $F_{io}$ is the time component of the $i$-meson field in
the outside region.

The energy associated to the density $\varepsilon(r)=T^0_0$, where
$T^\alpha_\beta$ is the energy-momentum tensor of the system, can be
calculated in the spherically symmetric metric by (see, e.g.,
\cite{Lee4})
\begin{equation}
  E_{t} = 4\pi \int {\rm e}^{(\nu+\lambda)/2} r^2 \varepsilon(r) d r.
\end{equation}

Thus, the total surface tension can be written as the sum of three
contributions,
\begin{equation} \label{stsi}
  \sigma_{t} = \sigma_N + \sigma_e + \sigma_C,
\end{equation}
where we have introduced the nuclear surface tension following the
method of BBP \cite{Baym1},
\begin{eqnarray}
  \sigma_{N} &=& \sum_{i=n,p,\sigma,\omega,\rho}
  {\rm e}^{(\nu_{core}+\lambda_{core})/2} \bigg\{ \int_{-\infty}^{a_i}
  [\epsilon_{i} (\bar{r}) - \epsilon_{ib}] d\bar{r} \nonumber\\
  & & +
  \int_{a_i}^{\infty} [\epsilon_{i}(\bar{r}) - \epsilon_{io}] d\bar{r}
  \bigg\}, \label{stng}
\end{eqnarray}
the electron surface tension
\begin{eqnarray}
  \sigma_e &=& {\rm e}^{(\nu_{core}+\lambda_{core})/2}
  \bigg\{ \int_{-\infty}^{a_e} [\epsilon_{e}(\bar{r}) - \epsilon_{eb}] d\bar{r} \nonumber\\
  & & \quad +
  \int_{a_e}^{\infty} [\epsilon_{e}(\bar{r}) - \epsilon_{eo}] d\bar{r}
  \bigg\}, \label{stesi}
\end{eqnarray}
and the surface tension for the electric field as
\begin{equation} \label{stcg}
  \sigma_{C} = {\rm e}^{(\nu_{core}+\lambda_{core})/2} \int_{-\infty}^{\infty} \epsilon_E (\bar{r})
  d\bar{r}.
\end{equation}
with $\epsilon_i (\bar{r})$ the energy density of the $i$ species of
fermion or meson fields, $\epsilon_{ib}$ is the energy density of
the $i$ species of fermion or meson fields in the bulk region,
$\epsilon_{io}$ is the energy density of the $i$ species of fermion
or meson fields in the outside region, and $\epsilon_E (\bar{r}) =
E^2 /(8\pi)$ is the electrostatic energy density. In the curved
spacetime equation (\ref{ssmetric}), the electric field is given by
(see, e.g., \cite{Belvedere1})
\begin{equation}
  E = {\rm e}^{-(\lambda_{core} + \nu_{core})/2} \frac{dV}{dr}.
\end{equation}

It is important to clarify how the values of $n_{io}$ and
$\epsilon_{io}$ are obtained. As we showed in
Ref.~\cite{Belvedere1}, the Einstein-Maxwell-Thomas-Fermi equations
have to be solved under the constraint of global charge neutrality
and not local charge neutrality, as in the traditional TOV-like
treatment. In the latter locally neutral configurations, the
continuity of total pressure leads to neutron stars with a crust
starting from nuclear density, where the clusterization of nucleons
starts to be preferred over the homogenous phase of the core, all
the way up to low densities in the surface. The region between
nuclear density and the neutron-drip density, $\rho_{drip}\approx
4.3\times 10^{11}$~g~cm$^{-3}$, is called the \emph{inner crust},
and at lower densities, $\rho<\rho_{drip}$, the \emph{outer crust}.
In this case the continuity of the pressure does not ensure the
continuity of the particle generalized chemical potentials. For
electrons it implies an inconsistency since the mismatching of the
electrochemical potential implies the existence of a Coulomb
potential energy, not accounted for in such a treatment (see, e.g.,
Ref.~\cite{RuedaPLB}).

In the globally neutral case, there is a different core-crust
boundary problem: the generalized fermion chemical potentials have
to match, at the end of the core-crust transition boundary layer,
their corresponding values at the base of the crust (outside
region); i.e., they must satisfy a condition of continuity (see
Ref.~\cite{Belvedere1} for details). It implies that the values of
$n_{io}$ and $\epsilon_{io}$ depend on the density at the base of
the crust under consideration.

We first consider below in Section \ref{sec:stns} the surface
tension of the system neglecting the presence of the crust. Then,
the more realistic case of a neutron star with a crust is considered
in Section \ref{sec:dne}. Configurations with only outer crust as
well as configurations with both inner and outer crust are studied.

Turning to the Coulomb energy, it is important to remark that, owing
to the small charge separation present in the system in the core
bulk region, we can assume that the electric field only exists in
the transition layer surface. Thus we can consider the electrostatic
energy as a surface property of the system, hence contributing to
the surface energy. This is a major difference between the present
system and an ordinary nucleus where the electrostatic energy is a
volume property.

The relation between the surface energy and Coulomb energy is very
important for a nucleus. As shown by Bohr and Wheeler \cite{Bohr1}
when the condition
\begin{equation} \label{bwc}
  E_{coul} > 2E_{sur}
\end{equation}
is satisfied, the nucleus becomes unstable against nuclear fission;
here $E_{coul}$ is the Coulomb energy of the nucleus and $E_{sur}$
is the surface energy of the nucleus. It is important to recall that
the idealized picture of the deformed nucleus of Bohr and Wheeler is
represented by two positively charged spheres joined by a nuclear
attraction neck. It is thus the interplay of the Coulomb and nuclear
surface energies that determines the lower energy state. Following
this argument one could think that, since we are treating here a
globally neutral system, such an instability mechanism is absent.
However, the condition (\ref{bwc}) can be also obtained by
requesting that a uniformly charged spheroid, constructed from an
axially symmetric deformation at constant volume of a uniformly
charged sphere, be energetically favorable. From a careful look at
the derivation of Eq.~(\ref{bwc})--see, e.g.,
Ref.\cite{reed2009}--it can be seen that this result follows from
the fact that Coulomb energy of the unperturbed system (the sphere)
depends on the radius as $E_{coul}\propto R^{-1}$. Such an inverse
radius dependence holds also in the case of a uniformly charged
shell, and also in the case of the globally neutral massive nuclear
density cores studied in Refs.~\cite{Rotondo2011,ruedaprc2011},
which fully reflect the properties of the system studied in this
work. We then expect that the Bohr-Wheeler condition of instability
against fission given by Eq.~(\ref{bwc}) applies also to our system.
Clearly such a condition is obtained keeping the system at nuclear
density and neglecting the the extra binding effect of gravity.

In thermodynamics, the surface tension is related to the mechanical
work needed to increase a surface area \cite{Gennes1},
\begin{equation} \label{stth}
  dW = \sigma dS,
\end{equation}
here $\sigma$ is the surface tension, $dS$ is the variation of the
surface area, and $dW$ is the mechanical work needed to increase the
surface area of the system. In this point of view, a system with a
positive surface tension has an attractive nature, and a system with
a negative surface tension has a repulsive nature.

Equations (\ref{stsi})-(\ref{stcg}) show that the surface tension
mainly depends on the fermion density and meson field profiles and
the energy densities of fermions and meson fields. The energy
density of the $i$-fermion species is given by
\begin{eqnarray}
  \epsilon_i (\bar{r})
  &=& \frac{1}{8\pi^2} \bigg\{ P_i^F \sqrt{(P_i^F)^2 + \tilde{m}_i^2} [2(P_i^F)^2 +
  \tilde{m}_i^2] \nonumber \\
  & & - \tilde{m}^4 \ln\frac{P_i^F + \sqrt{(P_i^F)^2 + \tilde{m}_i^2}}{\tilde{m}_i}
  \bigg\} \label{edf},
\end{eqnarray}
and the energy densities of the meson fields are (see, e.g.,
\cite{Lee4})
\begin{eqnarray} \label{edmfg}
  \epsilon_{\sigma}(\bar{r})&=& \frac{1}{2} {\rm e}^{-\lambda_{core}} \bigg( \frac{d\sigma}{d\bar{r}}
  \bigg)^2 + U(\sigma),\\
  \epsilon_{\omega}(\bar{r})&=& \frac{1}{2} {\rm e}^{-(\lambda_{core}+\nu_{core})}\bigg( \frac{d\omega}{d\bar{r}}
  \bigg)^2 + \frac{1}{2} {\rm e}^{-\nu_{core}} m_{\omega}^2 \omega^2,
\end{eqnarray}
\begin{eqnarray}
  \epsilon_{\rho}(\bar{r})&=& \frac{1}{2} {\rm e}^{-(\lambda_{core}+\nu_{core})} \bigg( \frac{d\rho}{d\bar{r}}
  \bigg)^2 + \frac{1}{2} {\rm e}^{-\nu_{core}} m_{\rho}^2 \rho^2,\\
  \epsilon_{E}(\bar{r})&=& {\rm e}^{-(\lambda_{core}+\nu_{core})} \frac{1}{8\pi} \bigg(
  \frac{dV}{d\bar{r}}
  \bigg)^2.
\end{eqnarray}

We can solve Eqs.~(\ref{eqcomg})-(\ref{eqng}) together with the
$\beta$ equilibrium (\ref{betaeq}) to obtain the fermion density and
meson field profiles. Following the similar method in
Ref.~\cite{Belvedere1}, this system of equations can be numerically
solved with appropriate conditions and approximations:
\begin{itemize}
  \item set a value for baryon number density of the bulk region $n_{bb} =
  n_{nb}+n_{pb}$; 
  \item in the bulk core region the electron
  density $n_{eb}$ is nearly equal to the proton density $n_{pb}$,
  i.e. $n_{pb} \simeq n_{eb}$; 
  \item set values for ${\rm e}^{\nu_{core}}$ and ${\rm
  e}^{-\lambda_{core}}$;
  \item the values of $n_{io}$ have to match their corresponding
  values at the edge of the crust. 
\end{itemize}

\section{Surface tension neglecting the presence of a crust}
\label{sec:stns}

\subsection{Surface tension at nuclear saturation density}
\label{sec:stnns}

\begin{figure}[h]
\includegraphics[width=0.75\columnwidth]{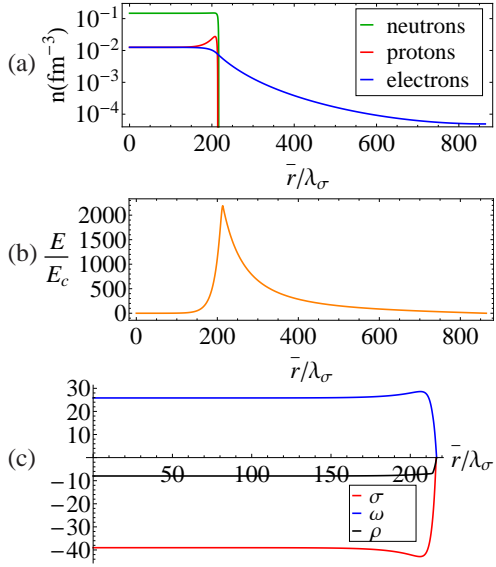}
\caption{(Color online) (a) Fermion density profiles in units of
${\rm fm}^{-3}$. (b) Electric field in units of the critical field
$E_c= m_e^2 c^3/(e\hbar)$. (c) Meson fields $\sigma$, $\omega$, and
$\rho$ in units of $\rm{MeV}$. Here $n_{bb} = n_{nucl}$, $n_{io}=0$,
and $({\rm e}^{\nu_{core}}, {\rm e}^{-\lambda_{core}}) \rightarrow
1$. $\lambda_{\sigma} = \hbar /(m_{\sigma}c) \sim 0.4$ ${\rm fm}$ is
the Compton wavelength of the $\sigma$ meson.} \label{SOREM}
\end{figure}

We first consider in this section the surface properties of this
transition layer neglecting the presence of the crust and the
gravitational interaction, i.e., $n_{io} = 0$ and $({\rm
e}^{\nu_{core}}, {\rm e}^{-\lambda_{core}}) \rightarrow 1$, as a
special case to gain some physical insight into this transition
layer. Also we assume here the baryon number density of the bulk
region to be the nuclear saturation density, $n_{bb} = n_{nb}+n_{pb}
= n_{nucl} = 0.16$ ${\rm fm}^{-3}$. The solution of
Eqs.~(\ref{eqcomg})-(\ref{eqng}) in this case is shown in
Fig.~\ref{SOREM}. Since the fermion densities tend to be zero in the
outside region, the thickness of the surface region for electrons
should be infinite. However, we just show the results up to a very
small electron density here, due to the plot scale and the accuracy
of the numerical calculation. As shown in Fig.~\ref{SOREM}, before a
sharp decrease of the proton and neutron densities, there is a bump
on the proton density profile due to Coulomb repulsion while the
electron density profile decreases.

Using the definitions in Eqs.~(\ref{stsi})-(\ref{stcg}), we can
calculate the surface tensions for this transition layer. The
results are shown in Table \ref{tablestns}. In order to study the
effect of the $\rho$ meson, we also show in Table \ref{tablestns}
the surface tensions in the case when the $\rho$ meson is not
present. The presence of $\rho$ decreases the total surface tension
$\sigma_{t}$ but increases the Coulomb energy, and so $\sigma_C$. We
can see that the difference of the surface tension for nucleons,
$\sigma_N$, in the presence and absence of the $\rho$ meson is
relatively small with respect to the changes on the electron
component and the electric field. We can explain this small
difference from the fact that, although the $\rho$ meson increases
the proton to neutron density ratio, in neutron stars the $\beta$
equilibrium in the presence of degenerate electrons leads to a high
isospin asymmetry $1-2 Z/A\approx 1$, hence the system is still
dominated by the neutron component, as we show below.

\begin{table}[h]
\begin{center}
\begin{tabular}{l|c|c|c|c}
\hline & $\sigma_{t}$ &$\sigma_N$ & $\sigma_e$ &
       $\sigma_C$\\
\hline  $\sigma$ $\omega$           & $6.28$     & $7.07$ & $-1.72$ & $0.92$\\
        $\sigma$ $\omega$ $\rho$    & $3.10$     & $7.30$ & $-8.34$ & $4.14$\\
\hline
\end{tabular}
\end{center}
\caption{Total and specific surface tensions in MeV fm$^{-2}$ of the
transition layer with and without the presence of the $\rho$ meson.
We set here $n_{bb} = n_{nucl}$, $n_{io}=0$, and $({\rm
e}^{\nu_{core}}, {\rm e}^{-\lambda_{core}}) \rightarrow 1$.}
\label{tablestns}
\end{table}

It is interesting to compare the above results with the nuclear
surface tension in the literature. The nuclear surface properties at
saturation density have been widely discussed in relativistic and
nonrelativistic models. The value of the nuclear surface tension in
the literature is around $\sigma_{L} \sim 1$ MeV fm$^{-2}$; see,
e.g., Refs.~\cite{Centelles1, Estal2}. The difference between our
result in Table \ref{tablestns} and $\sigma_{L}$ is mainly due to
the fact that the presence of degenerate electrons changes the
proton and neutron density profiles and also leads to a high isospin
asymmetry of our system as a byproduct of the $\beta$-equilibrium
condition. Especially, there is a bump on the density profiles in
our system, which would not appear in the case of normal nuclear
matter. Further discussions about this point is given below; see
also Table \ref{tablestnses}.

In order to understand where the surface tension comes from, we
calculate the contribution of each fermion and meson field to the
surface tension as
\begin{eqnarray}
  \sigma_n &=& \int_{-\infty}^{a_n} [\epsilon_{n}(z) - \epsilon_{nb}] dz +
  \int_{a_n}^{\infty} [\epsilon_{n}(z) - \epsilon_{no}] dz
  ,\label{sigman}\\
  \sigma_p &=& \int_{-\infty}^{a_p} [\epsilon_{p}(z) - \epsilon_{pb}] dz +
  \int_{a_p}^{\infty} [\epsilon_{p}(z) - \epsilon_{po}] dz
  ,\label{sigmap} \\
  \sigma_e &=& \int_{-\infty}^{a_e} [\epsilon_{e}(z) - \epsilon_{eb}] dz +
  \int_{a_e}^{\infty} [\epsilon_{e}(z) - \epsilon_{eo}] dz
  ,\label{sigmae}
\end{eqnarray}
\begin{eqnarray}
  \sigma_{\sigma} &=& \int_{-\infty}^{a_{\sigma}} [\epsilon_{\sigma}(z) - \epsilon_{\sigma b}] dz +
  \int_{a_{\sigma}}^{\infty} [\epsilon_{\sigma}(z) - \epsilon_{\sigma o}] dz
  ,\label{sigmasig}\\
  \sigma_{\omega} &=& \int_{-\infty}^{a_{\omega}} [\epsilon_{\omega}(z) - \epsilon_{\omega b}] dz +
  \int_{a_{\omega}}^{\infty} [\epsilon_{\omega}(z) - \epsilon_{\omega o}] dz
  ,\label{sigmaome}\\
  \sigma_{\rho} &=& \int_{-\infty}^{a_{\rho}} [\epsilon_{\rho}(z) -
  \epsilon_{\rho b}] dz +
  \int_{a_{\rho}}^{\infty} [\epsilon_{\rho}(z) - \epsilon_{\rho o}] dz . \label{sigmarho}
\end{eqnarray}
The results are shown in Table \ref{tablestnses}. For the sake of
comparison we also show the results in the case of ordinary nuclear
matter, namely for a system  without the presence of electrons. As
shown in Ref.~\cite{Belvedere1}, comparing to the profiles in the
case without the presence of the $\rho$ meson, the presence of the
$\rho$ meson leads to larger proton and electron densities, and a
larger bump of proton density happens. This effect is felt
indirectly by neutrons (although much less strongly), due to the
coupled nature of the system of equations
(\ref{eqcomg})-(\ref{eqng}). There is no such bump of the profiles
in the case of normal nuclear matter. Comparing the results of the
three cases in Table \ref{tablestnses}, the effect of the bump of
proton density on the surface tension is significant. The bump on
the profiles decreases the value of the surface tension for fermions
and increases the one for bosons. These results provide evidence of
large effect of electromagnetic interaction and electrons on the
proton and neutron density profiles, and therefore on the global
value of the surface energy of the system. It can be seen from Table
\ref{tablestnses} that we obtain a surface tension of ordinary
nuclear matter at saturation density (see the last line),
$\sigma_{N} \approx 1.7 $ MeV fm$^{-2}$. In our calculation, $n_n$
is slightly larger than $n_p$ according to the $\beta$ equilibrium.
This result is in agreement with the nuclear surface tension with a
small neutron excess, e.g., in Ref.~\cite{Centelles1}.

\begin{table}[h]
\begin{center}
\begin{tabular}{l|c|c|c|c|c|c}
\hline &$\sigma_{n}$ & $\sigma_p$ & $\sigma_e$ & $\sigma_{\sigma}$ &
$\sigma_{\omega}$
        & $\sigma_{\rho}$ \\
\hline  $n$ $p$ $e$ $\sigma$ $\omega$
        & $3.54$     & $-0.36$ & $-1.72$ & $3.16$ & $0.73$ &\\
        $n$ $p$ $e$ $\sigma$ $\omega$ $\rho$  & $-27.35$
        & $-5.19$ & $-8.34$ & $22.20$ & $19.93$ & $-2.28$\\
        $n$ $p$ $\sigma$ $\omega$ $\rho$ & $19.43$
        & $12.23$ &  & $-16.08$ & $-13.83$ & $-0.04$\\
\hline
\end{tabular}
\end{center}
\caption{Contribution of each fermion and meson field to the surface
tension, in MeV fm$^{-2}$. First row: the transition layer without
the presence of the $\rho$ meson. Second row: the transition layer
with the presence of the $\rho$ meson. Third row: normal nuclear
matter (without the presence of electrons). We set $n_{bb} =
n_{nucl}$, $n_{io}=0$, and $({\rm e}^{\nu_{core}}, {\rm
e}^{-\lambda_{core}}) \rightarrow 1$.} \label{tablestnses}
\end{table}

\subsection{Influence of baryon number density on the surface
tension} \label{sec:dbn}

\begin{figure}[h]
\includegraphics[width=0.85\columnwidth]{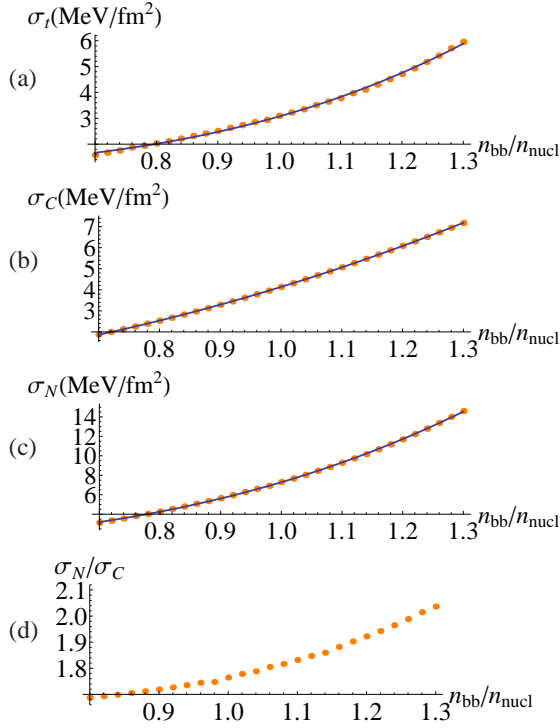}
\caption{(Color online) The dependence of the surface tension of the
transition layer on the baryon number density in the bulk region.
Here $n_{io}=0$ and $({\rm e}^{\nu_{core}}, {\rm
e}^{-\lambda_{core}}) \rightarrow 1$. (a) The total surface tension
$\sigma_{t}$, compared with the fit given in Eq. (\ref{fitsurfdnb}).
(b) Surface tension for the electric field $\sigma_C$, compared with
the fit given in Eq. (\ref{fitednb}). (c) Surface tension for
nucleons $\sigma_N$, compared with the fit given in Eq.
(\ref{fitndnb}). (d) Ratio of the surface tension for nucleons and
the surface tension for the electric field $\sigma_{N}/\sigma_{C}$.}
\label{stnbb}
\end{figure}

In order to study the dependence of the surface tension on the
baryon number density, we calculate the surface tensions for
different $n_{bb}$ following the similar procedure in
Sec.~\ref{sec:stnns}. The results are shown in Fig.~\ref{stnbb}.
Here the presence of the crust and the gravitational interaction is
neglected, i.e., $n_{io} = 0$ and $({\rm e}^{\nu_{core}}, {\rm
e}^{-\lambda_{core}}) \rightarrow 1$. From the results, the total
surface tension can be fitted by
\begin{equation} \label{fitsurfdnb}
  \sigma_{t,fit} = 1.05 + 2.02
  \bigg(\frac{n_{bb}}{n_{nucl}}\bigg)^{3.33}~\left({\rm MeV\,fm}^{-2}\right),
\end{equation}
the surface tension for the electric field can be fitted by
\begin{equation} \label{fitednb}
  \sigma_{C,fit} = -0.37 + 4.50
  \bigg(\frac{n_{bb}}{n_{nucl}}\bigg)^2~\left({\rm MeV\,fm}^{-2}\right),
\end{equation}
and the surface tension for nucleons can be fitted by
\begin{equation} \label{fitndnb}
  \sigma_{N,fit} = 0.95 + 6.33
  \bigg(\frac{n_{bb}}{n_{nucl}}\bigg)^{2.91}~\left({\rm MeV\,fm}^{-2}\right).
\end{equation}

As shown by BBP in \cite{Baym1}, the phenomenological surface
tension for nucleons within the Thomas-Fermi approximation can be
written as
\begin{equation} \label{stbbp}
  \sigma^{\rm BBP}_{sur} = B (W_o -W_i)^{\frac{1}{2}} (n_i -
  n_o)^{\frac{3}{2}},
\end{equation}
where $B$ is a constant, $W_o$ and $W_i$ are the binding energies
per nucleon in the outside and inside bulk regions, $n_o$ and $n_i$
are the nucleon number densities in the outside and inside bulk
regions. In the case of this section, we set the fermion densities
and meson fields to be zero in the outside region, i.e.,
$n_o=W_o=0$. Since the fractional concentration of protons in the
system we consider here is small, the binding energy per nucleon is
\cite{Baym1}
\begin{eqnarray}
  W(k,x)&=& W(k,0) + f(x) \nonumber\\
  &\approx& 19.74 k^2 - k^3 \frac{40.4 - 1.088 k^3}{1+ 2.545k} +
  f(x), \label{wkbbp}
\end{eqnarray}
where $k$ is defined by $n=2 k^3/(3\pi^2)$, with $n$ the nucleon
number density, and $x$ is the fractional concentration of protons.
The function $f(x)$ is a small correction to $W(k,0)$ since $x$ is
small in our system. From Eq.~(\ref{wkbbp}), one can estimate that
the leading term in the binding energy $W_i$ is the kinetic term,
proportional to $k^2$, i.e., $W_i\propto k^2 \propto n_{bb}^{2/3}$.
Thus one can estimate that $\sigma^{\rm BBP}_{sur} \propto
n_{bb}^{11/6}$ in the BBP phenomenological result \cite{Baym1},
where the effect of electromagnetic interaction on the profile of
fermion density is neglected. This BBP phenomenological result is
different from our result in Eq.~(\ref{fitndnb}). This is due to the
fact that the electromagnetic interaction and the presence of
electrons change the proton and neutron density profiles.

For $\sigma_C$, as shown in Eq.~(\ref{fitednb}) the surface tension
for the electric field is proportional to the square of the baryon
number density. This result can be understood as follows. The
Thomas-Fermi equilibrium condition for electrons given by
Eq.~(\ref{eqeg}) tells us that the Coulomb potential in the bulk
core is proportional to the bulk electron chemical potential, so
$V_b\propto \mu_{eb}$, and since the electrons are ultrarelativistic
at these densities we have $V_b\propto P^F_{eb}\propto
n_{eb}^{1/3}$. The thickness of the layer is of order $\Delta r\sim
n_{eb}^{-1/3}$ and so the electric field scales as $E\sim -\Delta
V/\Delta r \sim V_{b}/\Delta r\propto n_{eb}^{2/3}$. Thus the
contribution of the Coulomb energy to the surface tension satisfies
$\sigma_C \propto E^2 \Delta r\propto n_{eb}$ and since in the bulk
core we have $n_{eb}\simeq n_{pb}$ we obtain $\sigma_C \propto
n_{eb}=y n_{bb}$, where $y=n_{pb}/n_{bb}$ is the proton fraction in
the bulk region. In neutron stars the $\beta$ equilibrium between
neutrons, protons, and electrons leads to a highly nuclear isospin
asymmetry ($y\ll 1$), and since the nucleons are approximately
nonrelativistic and the electrons ultrarelativistic around nuclear
saturation density, it can be estimated from Eq.~(\ref{betaeq}) that
the proton fraction is proportional to the baryon density, i.e.,
$y\propto n_{bb}$, and therefore we finally obtain our final result
$\sigma_C \propto n^2_{bb}$.

In Fig.~\ref{stnbb} we show also the nuclear-to-Coulomb surface
tension ratio $\sigma_N/\sigma_C$. We find that this ratio is larger
than unity for all baryon number densities we considered. This would
in principle imply that the system is stable with respect to the
Bohr-Wheeler condition (\ref{bwc}) as we have previously discussed.

It is also worth mentioning that the result that $\sigma_N/\sigma_C
>1$ for every nucleon density in our system can be explained as the
result of the penetration of the relativistic electrons into the
nucleus (see Refs.~\cite{Rotondo2011,ruedaprc2011} for details).
This is allowed for configurations with sufficiently large sizes
$r_0 A^{1/3}>\hbar/(m_e c)$ or mass numbers $A > \hbar^3/(r_0 m_e
c)^3 \sim 10^7$, where $r_0\approx 1.2$ fm. For systems with much
larger mass numbers such as neutron stars, $A_{NS}\sim 10^{57}$, the
penetration of electrons is such that they nearly neutralize the
system and the electric field becomes appreciable only near the core
surface \cite{Rotondo2011,ruedaprc2011}.

However, the transition layer could be unbound if the gravitational
binding energy of the shell to the core is smaller than its
electrostatic energy. An approximate computation of the stability of
the transition layer in the above sense can be found in
Ref.~\cite{Rotondo2011}, where it was shown within Newtonian gravity
that the layer is gravitational bound provided the system has a
number of baryons $A\gtrsim 0.004(Z/A)^{1/2} (m_{Pl}/m_N)^3\sim
10^{55}(Z/A)^{1/2}$ or a mass $M=m_N A \gtrsim 0.01(Z/A)^{1/2}
M_\odot$, where $m_N$ and $m_{Pl}=(\hbar c/G)^{1/2}$ are the nucleon
and Planck masses. It is clear that this stability requirement
implies a lower limit for our globally neutral neutron stars. It
would be interesting to perform a detailed calculation taking into
account the effects of general relativity as well as of the magnetic
field on the transition surface induced by rotation (see
Ref.~\cite{kuantayrotondo}) and the centrifugal potential acting on
the shell. However, such calculation is out of the scope of this
work and will be presented elsewhere.

\section{Surface tension in presence of the crust}\label{sec:dne}

It was shown in Ref.~\cite{Belvedere1} that the properties of the
core-crust transition boundary layer depend on the nuclear
parameters, especially on the nuclear surface tension, and on the
density at the crust edge. The crust is composed of a nuclei lattice
in a background of degenerate electrons, whose density at the edge
of the crust is denoted here as $n_{e}^{crust}$. There are in
addition free neutrons in the crust when the density of the crust,
$\rho_{crust}$, is higher than the neutron-drip value
$\rho_{drip}\sim 4.3 \times 10^{11}$ g cm$^{-3}$ \cite{Baym1}. So
when the density of the crust $\rho_{crust}$ is smaller than the
neutron-drip value, i.e., $\rho_{crust} < \rho_{drip}$, we set the
proton and neutron densities to zero in the outside region while the
electron density must match the value $n_{e}^{crust}$, i.e., $n_{eo}
= n_{e}^{crust}$. In the cases when $\rho_{crust}
> \rho_{drip}$ both neutrons and electrons have to match their
corresponding crust values at the end of the core-crust transition
layer, i.e., $n_{eo} = n_{e}^{crust}$ and $n_{no} = n_{n}^{crust}$,
$n_n^{crust}$ being the neutron density at the crust edge.

As shown by BBP \cite{Baym1} there is no proton-drip at any density
of interest in these systems and therefore we keep zero as the
outside proton density value. In order to set the matching density
values for electrons and neutrons we use the relation of the free
neutron and electron densities in Section 6 of the work by BBP
\cite{Baym1}. At the neutron-drip point the electron Fermi momentum
is around $P^{F}_{eo}\approx 26$ MeV or $P^{F}_{eo}/P^{F}_{eb}
\approx 0.18$, where $P^F_{eo}$ is the electron Fermi momentum in
the outside region and $P^F_{eb}$ is the electron Fermi momentum in
the bulk region.

\begin{figure}[h]
\includegraphics[width=0.85\columnwidth]{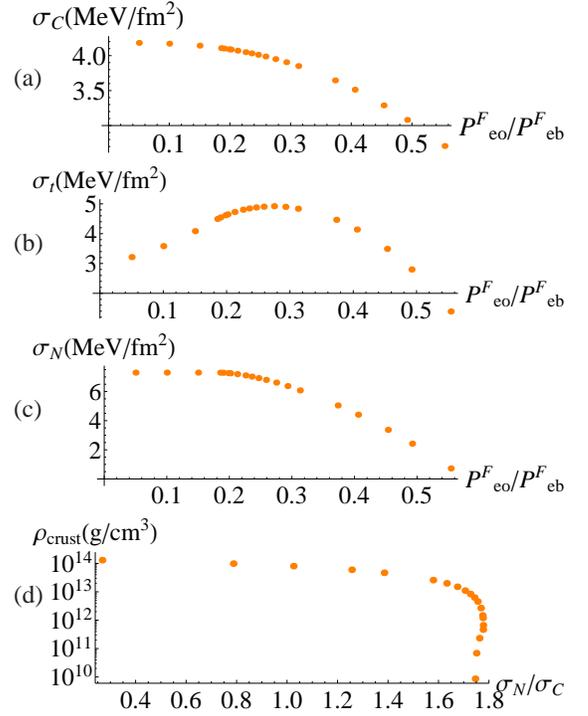}
\caption{(Color online) Dependence of the surface tension of the
transition layer on the fermion densities in the outside region and
the density of the crust. Here $n_{bb} = n_{nucl}$ and $({\rm
e}^{\nu_{core}}, {\rm e}^{-\lambda_{core}}) \rightarrow 1$. (a)
Surface tension for the electric field, $\sigma_C$. (b) The total
surface tension $\sigma_{t}$. (c) The surface tension for nucleons,
$\sigma_{N}$. (d) Ratio of the surface tension for nucleons and the
surface tension for the electric field, $\sigma_{N}/\sigma_{C}$,
with respect to the density of the crust $\rho_{crust}$. The neutron
drip point $\rho_{drip}\sim 4.3 \times 10^{11}$ g cm$^{-3}$ is
around $P^{F}_{eo}/P^{F}_{eb} \approx 0.18$.} \label{stneo}
\end{figure}

The results of the dependence of the surface tension on the outside
electron densities and the density of the crust are shown in
Fig.~\ref{stneo}. Here we also neglect the presence of the
gravitational interaction, i.e., $({\rm e}^{\nu_{core}}, {\rm
e}^{-\lambda_{core}}) \rightarrow 1$.

The results of Fig.~\ref{stneo} show that the Bohr-Wheeler condition
(\ref{bwc}) for the instability is reached at a crust density
$\rho_{crust}^{crit} \sim $ $1.2 \times 10^{14}$ g cm$^{-3}$, so the
system becomes unstable against fission when
$\rho_{crust}>\rho_{crust}^{crit}$, imposing a physical upper limit
to the density at the edge of the crust. It becomes interesting to
include the binding effect of gravity and any other attractive
contribution that strengthens the stability of the system, which
will be analyzed elsewhere. It is interesting that this upper limit
on the crust density implies a lower limit to the maximum electric
field in the core-crust transition region, limiting at the same time
to approaching a state of quasilocal charge neutrality of the
neutron star.

As shown in Fig.~\ref{stneo}, the surface tension for the electric
field decreases as increasing the electron number density in the
outside region. The reason is that the increasing electron number
density in the outside region \cite{Belvedere1} causes a decrease of
the thickness of the interface and of the proton and electron
density difference; i.e., the surface charge density decreases.

It is shown in Fig.~\ref{stneo} that the dependence of the surface
tension for nucleons, $\sigma_{N}$, on the electron number density
in the outside region is weak before the neutron drip point. The
influence of electron density in the outside region on the surface
structure of nucleons is small in this case. After the neutron drip
point, the free neutrons in the outside region lower the surface
tension significantly, as expected in the BBP phenomenological
result \cite{Baym1}. In addition, as shown in Fig.~\ref{stneo}, the
total surface tension $\sigma_{t}$ first increases and then
decreases with increasing fermion densities in the outside region.
This is due to the combination of the following two effects. (I) as
shown in Table \ref{tablestnses}, the contribution of electrons to
the total surface tension is negative. For increasing electron
density in the outside region, the effect of electrons on the
surface tension becomes weaker. This increases the total surface
tension. (II) After the neutron drip point, the surface tension for
nucleons $\sigma_{N}$ is lowered significantly by the free neutrons
in the outside region.

\section{Effects of the gravitational
interaction on the surface tension} \label{sec:stwg}

We turn now to analyze the effects of the inclusion of the
gravitational field on the surface tension of this transition layer.
For the sake of simplicity, we make this analysis in the simplest
case without a crust, considered in Section \ref{sec:stns}.

\begin{figure}[h]
\includegraphics[width=0.75\columnwidth]{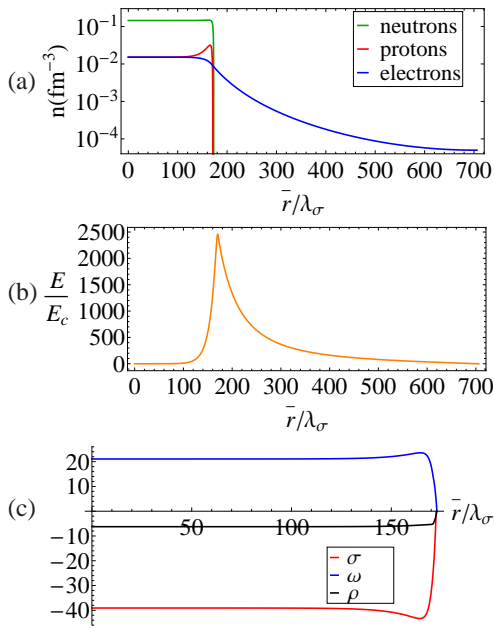}
\caption{(Color online) (a) Fermion density profiles in units of
$\rm{fm^{-3}}$. (b) Electric field in units of the critical field
$E_c$. (c) Meson fields $\sigma$, $\omega$, and $\rho$ in units of
$\rm{MeV}$. Here we set ${\rm e}^{\lambda_{core}} \approx {\rm
e}^{-\nu_{core}} = 1.5$, $n_{bb} = n_{nucl}$, and $n_{io}=0$.}
\label{elam15}
\end{figure}

As shown in Ref.~\cite{Belvedere1}, at the core radius (in this case
the surface) of the neutron star, the metric functions are
approximately the same as the Schwarzschild solution, so at the
border of the star we have
\begin{equation} \label{metriccore}
  {\rm e}^{\nu_{core}} \approx {\rm e}^{-\lambda_{core}} = 1 -
  \frac{2GM(R_{core})}{R_{core}},
\end{equation}
with $M(R_{core})$ the mass of the star. The results of the solution
of Eqs.~(\ref{eqcomg})-(\ref{eqng}) are shown in Fig.~\ref{elam15}
for the case ${\rm e}^{\lambda_{core}} \approx {\rm e}^{-\nu_{core}}
= 1.5$.

Comparing to the results shown in Fig.~\ref{SOREM}, the fermion
density and meson field profiles are similar to their counterparts
in the case without the gravitational field. In Fig.~\ref{elam15} we
see a larger proton density, a smaller neutron density, and a
smaller size of the core-crust transition layer leading to a larger
maximum of the electric field, comparing to Fig.~\ref{SOREM}.

\begin{figure}[h]
\includegraphics[width=0.85\columnwidth]{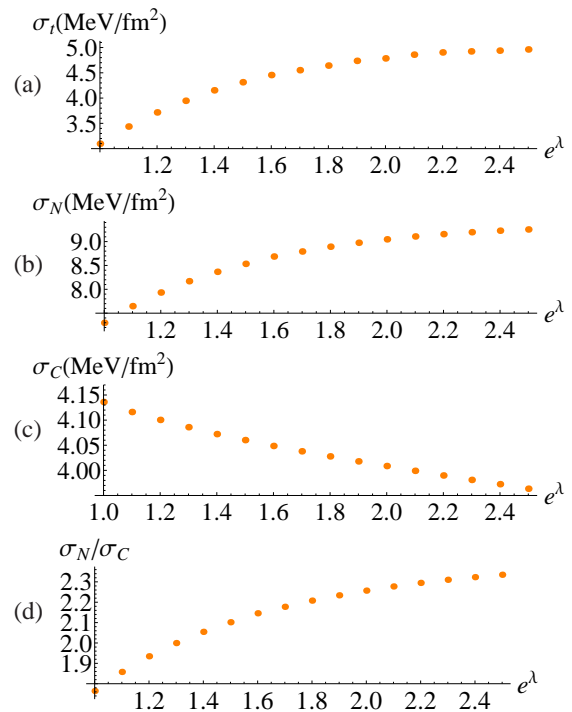}
\caption{(Color online) The dependence of the surface tension of the
transition layer on the value of metric ${\rm e}^{\lambda_{core}}$.
(a) The total surface tension $\sigma_{t}$. (b) Surface tension for
nucleons, $\sigma_{N}$. (c) Surface tension for the electric field,
$\sigma_{C}$. (d) Ratio of the surface tension for nucleons and the
surface tension for the electric field, $\sigma_{N}/\sigma_{C}$. We
set here $n_{bb} = n_{nucl}$ and $n_{io}=0$.} \label{stdelam}
\end{figure}

Figure \ref{stdelam} shows the results of the dependence of the
surface tension on the value of metric ${\rm e}^{\lambda_{core}}$.
As shown in Fig.~\ref{stdelam}, the total surface tension and the
surface tension for nucleons increase as increasing the value of the
metric ${\rm e}^{\lambda_{core}}$. There are two effects which
influence the characters of the total surface tension and the
surface tension for nucleons. First, as we have seen the presence of
gravitational field changes the fermion density and meson field
profiles. Second, the difference between the proton density and the
neutron density becomes smaller when the value of the metric ${\rm
e}^{\lambda_{core}}$ increases, lowering the isospin asymmetry of
the system. The combination of these two effects leads to the
characters of the total surface tension and the surface tension for
nucleons shown in Fig.~\ref{stdelam}. In addition, as shown in
Fig.~\ref{stdelam}, the change of the value of the surface tension
for the electric field when increasing the value of ${\rm
e}^{\lambda_{core}}$ is small. That is due to the balance of the
following two effects: (I) the electric field in the surface region
becomes larger (see Fig.~\ref{elam15}); (II) the thickness of the
surface becomes smaller, and then the Coulomb energy distributes in
a smaller region. It can be also checked from Fig.~\ref{elam15} how
in the limit ${\rm e}^{\lambda_{core}}\to 1$ all quantities tend to
the values found in Sec.~\ref{sec:stns} in the flat case.

\section{Summary and discussion}
\label{sec:sum}

Taking into account strong, weak, electromagnetic, and gravitational
interactions, and fulfilling the global charge neutrality of the
system, a transition layer will happen between the core and crust of
neutron stars \cite{Belvedere1}. This is different from the results
from traditional TOV equations imposing local charge neutrality.
This core-crust transition layer happens at the saturation density
of nuclear matter. In this article, using RMFT together with the
Thomas-Fermi approximation, we study the surface properties of this
transition layer. In particular, we computed the surface tension and
Coulomb energy of the transition shell and analyzed the role of each
fermion component and meson fields in the determination of the
properties of this core-crust transition layer.

Since the length scale of the core-crust transition layer ($\sim
\lambda_{e}$) is much smaller than the radius of neutron stars and
the electron density is nearly equal to the proton density in the
bulk hadronic phase of neutron star cores, we applied the
semi-infinite matter model as an approximation to construct the
surface tension for this core-crust transition layer, following the
method of BBP in Ref.~\cite{Baym1}. We first presented the studies
of this transition layer neglecting the presence of the
gravitational interaction. We calculated the surface tension and the
Coulomb energy for the transition layer of this system for different
baryon number densities near the nuclear saturation density. The
results show that the total surface tension as well as the surface
tension for the electric field and the surface tension for nucleons
are proportional to some power-law function of the baryon number
density in the bulk region; see
Eqs.~(\ref{fitsurfdnb})-(\ref{fitndnb}). The difference between the
surface energy of this neutron star matter and the phenomenological
results \cite{Baym1} in nuclear physics has been analyzed. We also
studied the surface structure for different fermion densities in the
outside region, namely for different densities of the neutron star
crust.

We also presented the analysis of the influence the gravitational
field and on the structure of the transition layer and the surface
tension. The results show that the fermion density and meson field
profiles are similar to the case without the presence of
gravitational field, although some quantitative differences appear.
We show that the total surface tension and the surface tension for
nucleons increase with increasing value of the metric function ${\rm
e}^{\lambda_{core}}$.

We studied the instability against Bohr-Wheeler surface deformation
for all the systems. We find that the instability sets in at a
critical density of the crust $\rho_{crust}^{crit} \sim $ $1.2
\times 10^{14}$ g cm$^{-3}$. This implies a lower limit to the
maximum electric field of the core-crust transition region and makes
inaccessible a state of quasilocal charge neutrality for the neutron
star, which will in principle be reached when the limit
$\rho_{crust}=\rho_{core}\approx \rho_{nucl}$ is approached.

The results of this work open the way to more general studies
relevant for the analysis of the stability of neutron stars and the
core-crust transition surface. Some of the effects that need to be
addressed for the stability of the shell include gravitational
binding, centrifugal repulsion, magnetic field induced by rotating
electric field and hence magnetic dipole-dipole interactions. It
would be interesting to perform a similar analysis for the case of
strange stars both bare and in the presence of an outer crust.

It is also important to mention that surface effects and boundary
layers are contained in the widely discussed nuclear pasta phases
(see, e.g., Refs.~\cite{Maruyama1,Avancini1,Okamoto1,Grill1}, and
references therein) expected in the low-density nuclear matter
composing the inner crust of neutron stars. Those configurations
also fulfill the condition of global charge neutrality. However, in
there the condition of global charge neutrality is only imposed in
the pasta phase while keeping the condition of local charge
neutrality in the rest of the configuration, e.g., in the core of
the neutron star. In contrast, in our model, the global charge
neutrality is fulfilled in the whole configuration, which leads to
the phenomenon of gravito-polarization in the core of the neutron
star. Along this line, it would be interesting to study the
differences of these two scenarios and to establish which is the
configuration of minimum energy and therefore realized in nature.
This is a very interesting question which deserves a detailed and
deep analysis; however, it is out of the scope of the present work
and we therefore leave it for a future publication.

To end, it is interesting to briefly discuss some of the observables
which could shed light into the structure of the neutron star and
therefore to probe the underlying theory.

On one hand, there might be some effects coming from the microscopic
structure. One possibility could be some electromagnetic processes
due to the strong electric field in the core-crust interface, such
as the annihilation line of $e^{-}e^{+}$ to two photons. These
$e^{-}e^{+}$ pairs can be produced by neutron star perturbations.
However, this effect could be difficult to observe with the current
instrumentation; we are planning to analyze in detail this
interesting problem elsewhere.

On the other hand, as we have pointed out, from the macroscopic
structure point of view the new structure of the neutron star leads
to different radii due to the different size of the crust. This
necessarily leads to the possibility of probing the theory of
neutron stars and in particular the physics of the core-crust
transition from reliable observations of their masses and radii.
Such measurements can come for instance from observations of the
thermal evolution of accreting and isolated neutron stars. In
particular, observations of the cooling of the neutron star during
its thermal relaxation phase ($t\lesssim 50$~yr after birth), where
the core and the crust are thermally decoupled, carry crucial
information on the core-crust transition density and therefore on
the crust mass and size \cite{inprep}.

If we move on to the last stages of the life of a neutron star, it
is clear that the electromagnetic structure of the neutron star is
particularly relevant for the process of its gravitational collapse.
A core endowed with electromagnetic structure leads to signatures
and energetics markedly different from the ones of a core endowed
uniquely of gravitational interactions \emph{\`a la} Oppenheimer and
Snyder \cite{OS39}; see, e.g.,
Refs.~\cite{vitagliano03a,vitagliano03b,ruffinireview08,physrep}. As
pointed out recently \cite{xue2012,xue2013}, in these cores there
are electric processes that might lead to a vast $e^{-}e^{+}$
production in the process of collapse to a black hole.

\vspace{0.5cm}
\noindent %
{\bf Acknowledgement}

Yuan-Bin Wu is supported by the Erasmus Mundus Joint Doctorate
Program by Grant Number 2011-1640 from the EACEA of the European
Commission.



\end{document}